\documentclass[12pt,fleqn]{iopart}
\usepackage{graphicx}
\usepackage{subfigure}
\begin{document}
\title{Strong-coupling study of spin-1 bosons in square and triangular optical lattices}

\author{Takashi Kimura}

\address{Department of Mathematics and Physics, Kanagawa University, 
2946 Hiratsuka, Kanagawa 259-1293, Japan}
\ead{tkimura@kanagawa-u.ac.jp}
\begin{abstract}
We examine the superfluid-Mott insulator (SF-MI) transition 
of antiferromagnetically interacting spin-1 bosons trapped 
in a square or triangular optical lattice. 
We perform a strong-coupling expansion up to the third 
order in the transfer integral $t$ between the nearest-neighbor lattices. 
As expected from previous studies, an MI phase with 
an even number of bosons is considerably 
more stable against the SF phase than it is with an odd number of bosons.
Results for the triangular lattice are similar to those for the 
square lattice, which suggests that  the lattice geometry 
does not strongly affect the stability of the MI phase against the 
SF phase. 
\end{abstract}

\maketitle

\section{Introduction}
The development of optical lattice systems based on laser technology
has renewed interest in strongly correlated lattice systems. 
One of the most striking phenomena of the optical-lattice systems
is the superfluid-Mott insulator (SF-MI) phase transition; 
the SF phase (i.e., the coherent-matter-wave phase) emerges when 
the kinetic energy is larger enough 
compared with the on-site repulsive interaction. Otherwise,  
the MI phase, i.e., the number-state phase without coherence emerges. 
The low-lying excitations of these optical-lattice systems 
can be described by using the Bose--Hubbard model. 
The temperature of trapped-atom systems can be extremely low, and 
hence, we hereafter assume the ground states of the system. 

Spin degrees of freedom also play an important role in optical-lattice 
systems. In theory, lots of analytical 
and numerical studies have been performed for the spin-1 Bose--Hubbard model
\cite{Review}, including 
rigorous results for a finite system \cite{exact}. 
In the case of antiferromagnetic spin-spin interactions, 
the perturbative mean-field approximation (PMFA) \cite{PMFA} 
indicates that 
when filling with an even number of bosons, 
the MI phase  is considerably more 
stable against the SF phase than when  filling with 
an odd number of bosons. 
This conjecture has been 
found by density matrix renormalization group (DMRG) \cite{DMRG} 
and quantum Monte Carlo (QMC) methods \cite{QMC1D1,QMC1D2} 
in one dimension (1D). Recently, QMC methods also confirmed 
that conjecture in a two-dimensional (2D) square lattice \cite{QMC2D}. 

Another interesting property of the spin-1 Bose--Hubbard model 
with antiferromagnetic interactions  
is the first-order phase transition: the SF-MI phase 
transition is of the first order in a part of the SF-MI phase diagram. 
The first-order transition has also been studied by using the 
variational Monte Carlo \cite{VMC} and QMC \cite{QMC2D} methods
in a 2D square lattice. The QMC results indicate that 
the phase transition can be of the first order, which 
is consistent with mean-field (MF) analysis \cite{First,F2,F3}. 
However, the first-order transition disappears for strong 
antiferromagnetic interactions; 
a MF calculation similar to that of Ref. \cite{First}  
and the QMC study \cite{QMC2D} 
show that the first-order SF-MI transition 
from the Mott lobe with two bosons per site disappears 
when $U_2/U_0 \ge 0.32$ and $U_2/U_0 \ge 0.15$, respectively.
Thus, we assume strong interactions where the SF-MI transition 
is of the second order. 

For the second-order SF-MI transition, 
the strong-coupling expansion of kinetic energy \cite{Freericks}
is excellent for obtaining the phase boundary. 
This method has been applied to 
the spinless \cite{Freericks,Freericks2}, 
extended \cite{Iskin}, hardcore \cite{Hen}, 
and two-species models \cite{Iskin2},   
and the results agree well 
with QMC results \cite{Freericks2,Hen}.  
Thus, in this study, we perform the strong-coupling expansion 
with the spin-1 Bose--Hubbard model. 
In another publication \cite{Strong}, we  examined 
the case of hypercubic lattices. In this study, 
we examine the triangular lattice and 
compare the results with those of a square lattice  
to clarify whether the lattice structure 
plays a key role for the SF-MI transition. 
The triangular lattice is intriguing because it 
frustrates the spin systems or spinful Fermi systems. 

The rest of this paper is organized as follows: 
Section II briefly introduces the spin-1 Bose--Hubbard model 
and the strong-coupling expansion. 
Section III provides our results. 
A summary of the results is given in Section IV. 
Some long equations that result from the strong-coupling 
expansion are summarized in  Appendix A. 
\section{Spin-1 Bose--Hubbard model and strong coupling expansion}
The spin-1 Bose--Hubbard model is given by 
$H=H_0+H_1$, where
\begin{eqnarray}
H_0&=&-t\sum_{\langle i,j\rangle,\alpha}(a_{i \alpha}^\dagger a_{j \alpha}^{}
 + a_{j \alpha}^\dagger a_{i \alpha}^{}), \nonumber\\
H_1&=&-\mu\sum_{i,\alpha} a_{i \alpha}^\dagger a_{i \alpha}^{} 
+\frac{1}{2}U_0\sum_{i,\alpha,\beta} a_{i \alpha}^\dagger a_{i \beta}^\dagger
a_{i \beta}^{}a_{i \alpha}^{}\nonumber\\
&&+\frac{1}{2} U_2 \sum_{i,\alpha,\beta,\gamma,\delta} 
a_{i \alpha}^\dagger a_{i \gamma}^\dagger
{\bf F}_{\alpha \beta} \cdot {\bf F}_{\gamma \delta} a_{i \delta}^{}
 a_{i \beta}^{}.\label{H}\nonumber\\
&=&\sum_i\Big[-\mu\hat{n}_i+\frac{1}{2}U_0\hat{n}_i(\hat{n}_i-1)
+\frac{1}{2}U_2({\hat{\mathbf S}_i}^2-2\hat{n}_i)\Big].
\end{eqnarray}
Here,  $\mu$ and $t(>\ 0)$ are the chemical potential 
and the hopping matrix element, respectively.  
$U_0$ ($U_2$) is the spin-independent (spin-dependent) 
interaction between bosons. We assume  repulsive ($U_0>0$) 
and antiferromagnetic ($U_2>0$) interaction. 
$a_{i \alpha}$ ($a_{i \alpha}^\dagger$) 
annihilates (creates) 
 a boson at site $i$ with spin-magnetic quantum number 
$\alpha=1,0,-1$. 
$\hat{n}_i\equiv\sum_\alpha n_{i\alpha}$ ($n_{i\alpha}\equiv 
a_{i \alpha}^\dagger a_{i \alpha}$)
is the number operator at site $i$. 
$ {\hat{\bf S}_i}\equiv \sum_{\alpha,\beta}a_{i \alpha}^\dagger 
{\bf F}_{\alpha \beta} a_{i \beta}$  
is the spin operator at site $i$, where 
${\bf F}_{\alpha \beta}$ represents the spin-1 matrices. 
In this study, we assume a tight-binding model 
with only nearest-neighbor hopping and  
$\langle i,j \rangle$ expresses sets of adjacent sites $i$ and 
$j$. 

When $t\rightarrow 0$, the ground state is the MI phase with the lowest
interaction energy. The number $n_0$ of bosons per site is odd 
when  $U_0(n_0-1)<\mu<U_0n_0-2U_2$, whereas it is even 
when $U_0(n_0-1)-2U_2<\mu<U_0n_0$.  The MI phase with an even 
number of bosons is 
\begin{equation}
\Psi_{\rm even}=\prod_k |n_0,0,0\rangle_k. \label{Me}
\end{equation}
Here,  $|n_0,0,0\rangle_k$ 
implies the boson number $n_0$, the spin magnitude 
$S=0$, and the spin magnetic quantum number $S_z=0$ at site $k$. 
However, for the MI phase with 
an odd number of bosons per site,  
we define a nematic state with $S_z=0$:  
\begin{equation} 
\Psi_{\rm odd}=\prod_k |n_0,1,0\rangle_k   \label{Mo}
\end{equation}
because we assume antiferromagnetic interactions. 
The dimerized state is degenerate with $\Psi_{\rm odd}$ 
for $t=0$ and is considered to be the ground state for finite $t$
in 1D. Therefore, the results based on $\Psi_{\rm odd}$ 
are basically limited to 2D or larger dimensions. 
%$\Psi_{\rm even}$ and $\Psi_{\rm odd}$ is also adopted as the ground state in PMFA \cite{Tsuchiya}, which we compare with the results in the following section. 

Next, we define the defect states by doping an extra particle or hole into 
$\Psi_{\rm even}$ and $\Psi_{\rm odd}$ as follows: 
\begin{eqnarray}
\Psi^{\rm part}_{\rm even}&=&
\frac{1}{\sqrt{N}} \sum_{i}\Big[
%f^{\rm part}_{{\rm even},i}
|n_0+1,1,0\rangle_i\otimes\prod_{k\ne i} |n_0,0,0\rangle_k\Big],\label{pe}\\
\Psi^{\rm hole}_{\rm even}&=&\frac{1}{\sqrt{N}} \sum_{i}\Big[
%f^{\rm hole}_{{\rm even},i}
|n_0-1,1,0\rangle_i
\otimes\prod_{k\ne i} |n_0,0,0\rangle_k\Big],\label{he}\\
\Psi^{\rm part}_{\rm odd}&=&\frac{1}{\sqrt{N}} \sum_{i}\Big[
%f^{\rm part}_{{\rm odd},i}
|n_0+1,0,0\rangle_i
\otimes\prod_{k\ne i} |n_0,1,0\rangle_k\Big],\label{po}\\
\Psi^{\rm hole}_{\rm odd}&=&\frac{1}{\sqrt{N}} \sum_{i}\Big[
%f^{\rm hole}_{{\rm odd},i}
|n_0-1,0,0\rangle_i
\otimes\prod_{k\ne i} |n_0,1,0\rangle_k\Big]. \label{ho}
\end{eqnarray} 
Here, $N$ is the number of lattice sites. 
We assume that these defect states can be regarded as the SF states 
doped with infinitesimal numbers of particles or holes. 
By applying the Rayleigh--Schr{\"o}dinger perturbation theory 
to these MI and defect states, we obtain the energy
of the MI state and that of the defect states up to the third order in $t$.  

\section{Results} 
The results for the energy per site of the MI state and of defect states
in the square lattice are summarized in Appendix A. 
The phase can be determined by these energies. 
Specifically, if $E_{\rm MI}(n_0,\mu,t)>(<) \min\Big(E^{\rm part}_{\rm def}(n_0,\mu,t),E^{\rm hole}_{\rm def}(n_0,\mu,t)\Big)$, then 
the phase is SF (MI),  
where $E_{\rm MI}(n_0,\mu,t)$ is the energy of the MI state 
and $E^{\rm part}_{\rm def}(n_0,\mu,t)$ [$E^{\rm hole}_{\rm def}(n_0,\mu,t)$] 
is the energy of the defect state with one extra particle (hole).
The SF--MI phase boundary is thus determined by
\begin{eqnarray}
  E_{\rm MI}(n_0,\mu,t)&=&E^{\rm part}_{\rm def}(n_0,\mu,t)
\end{eqnarray}
or
\begin{eqnarray}
  E_{\rm MI}(n_0,\mu,t)&=&E^{\rm hole}_{\rm def}(n_0,\mu,t). 
\end{eqnarray}

\begin{figure}[htdp]
\includegraphics[height=3in]{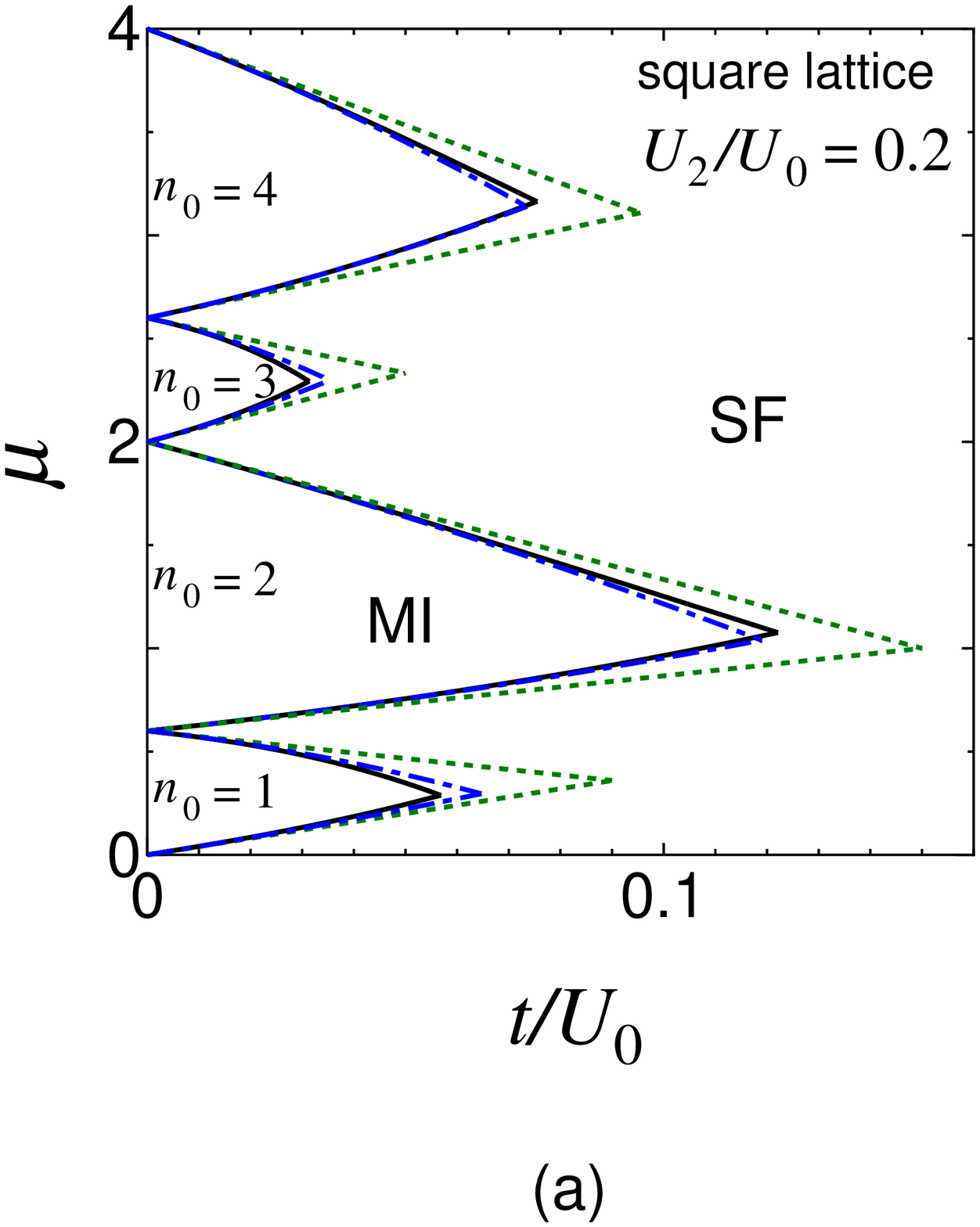}
\hspace{0.5in}
\includegraphics[height=3in]{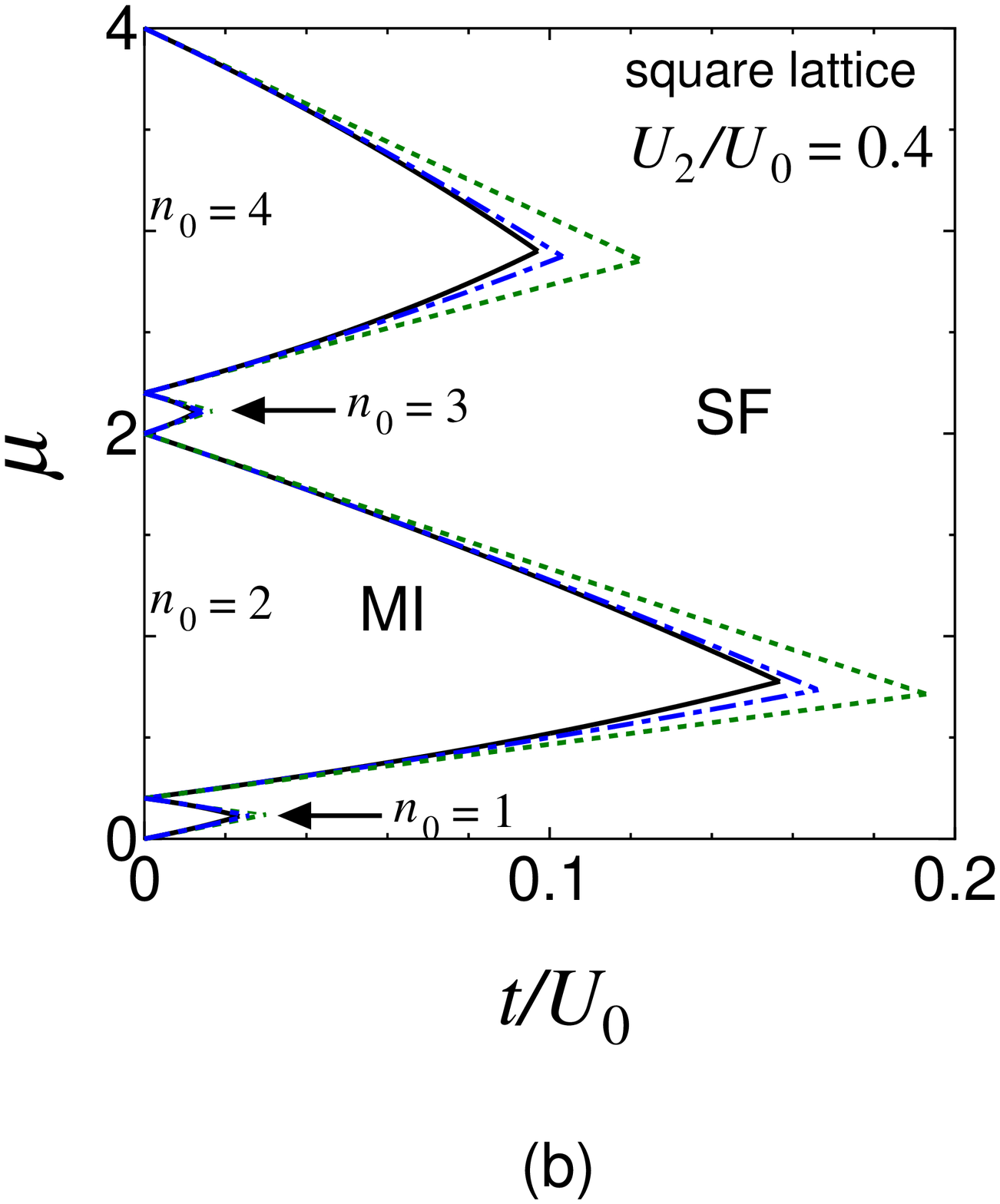}
\ \\ 
\caption{
(Color online) Phase diagrams  
obtained by strong-coupling expansion 
for (a) $U_2/U_0 = 0.2$ 
and for (b) $U_2/U_0 = 0.4$ 
in the square lattice. 
The solid curves show the results up to third order in t. 
Results up to first order (second order) in $t$ are also shown 
by the green dashed (blue dot-dashed) curve. 
}
\label{ex.sq}
\end{figure}
\begin{figure}[htdp]
 \includegraphics[height=3in]{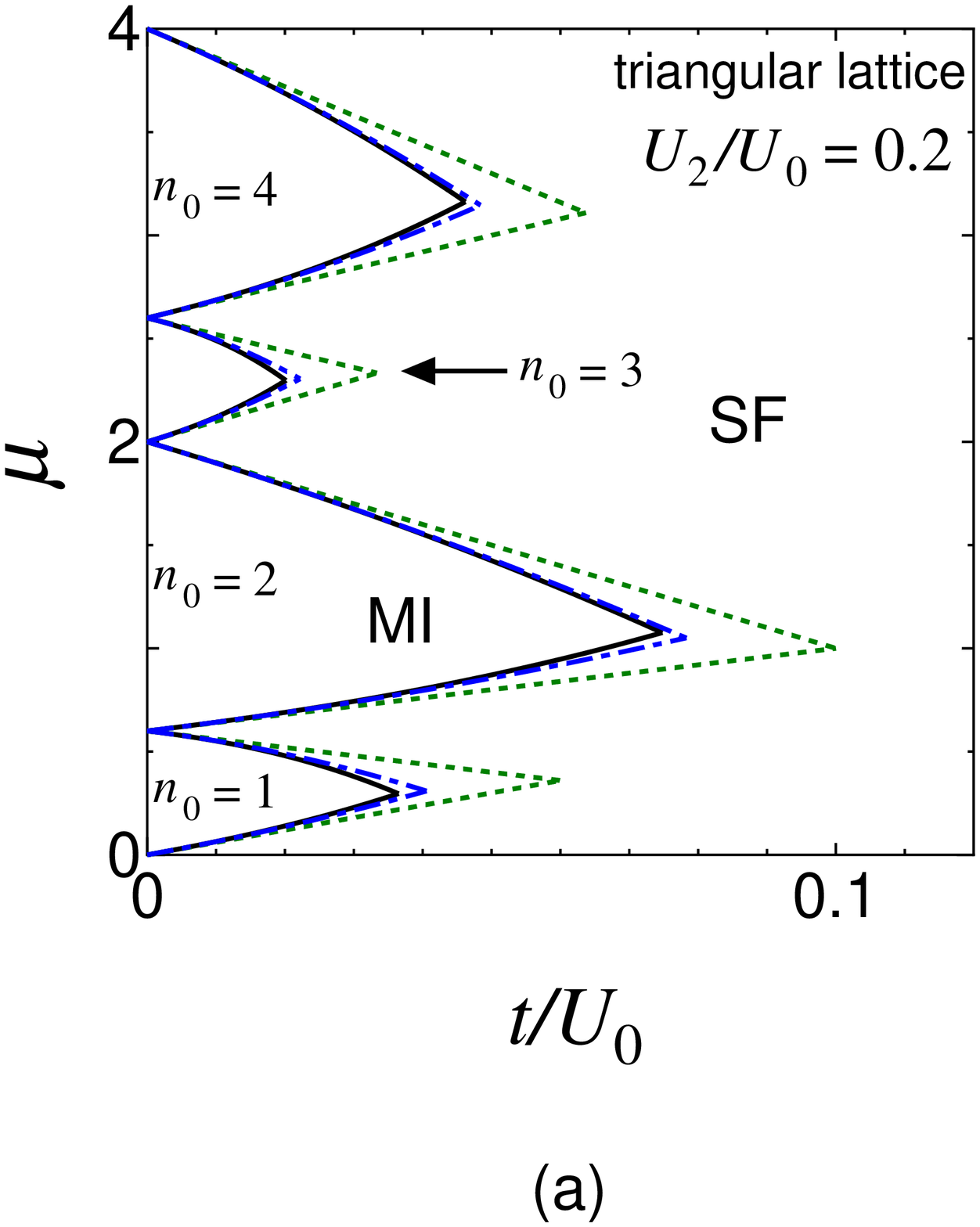}
\hspace{0.5in}
 \includegraphics[height=3in]{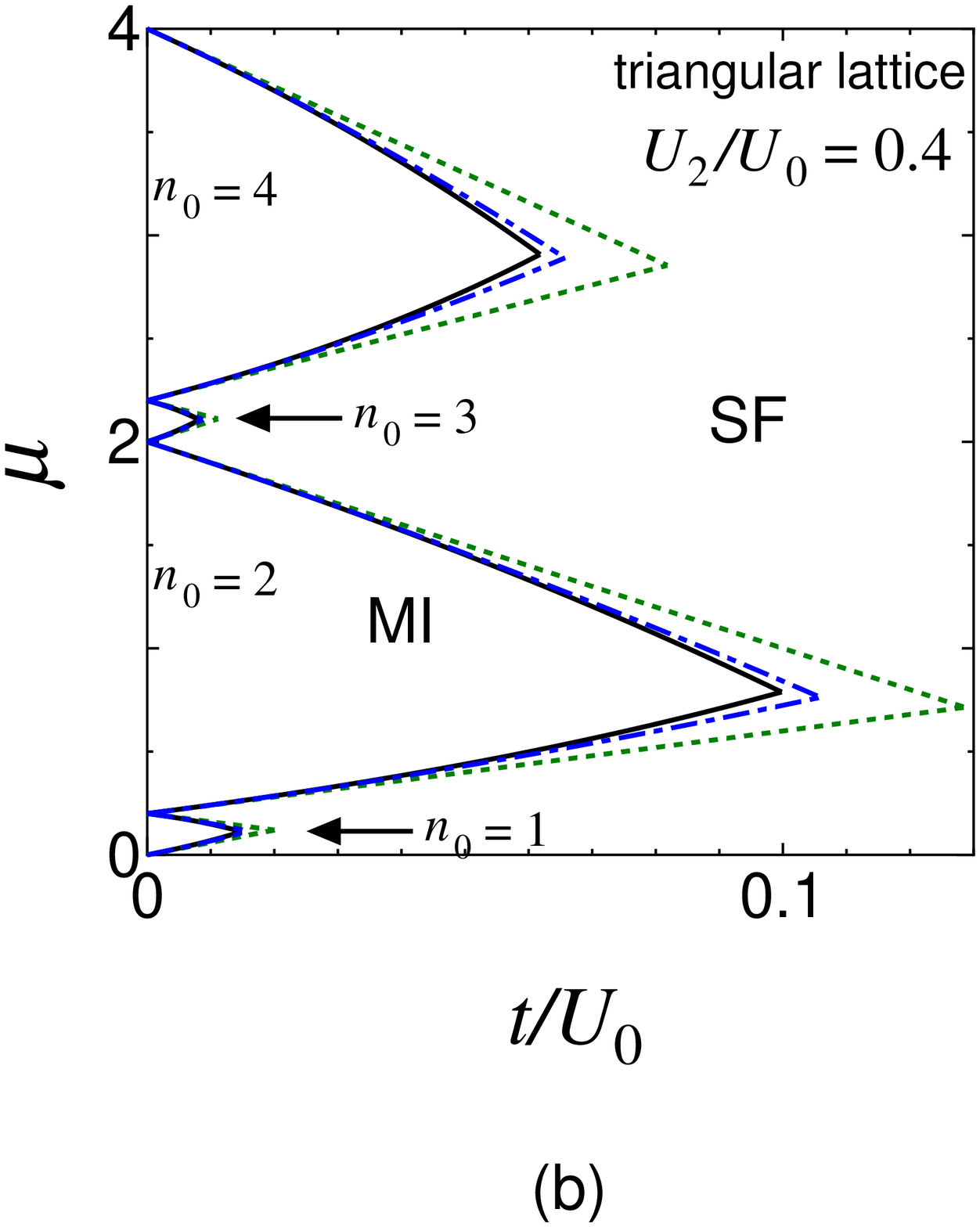}
\ \\ 
\caption{
(Color online) Same plot as in Fig. \ref{ex.sq}
but for (a) $U_2/U_0 = 0.2$ and (b) $U_2/U_0 = 0.4$ 
in the triangular lattice. 
}
\label{ex.tr}
\end{figure}

Figures \ref{ex.sq} and \ref{ex.tr}
show the phase diagram obtained from the strong-coupling expansion 
for the square lattice and for the triangular lattice, respectively. 
In both lattices, 
the MI phase for even-boson filling is considerably more stable 
against the SF phase than for odd-boson filling, which 
is consistent with QMC studies for the square lattice and MF studies. 
The area of the MI phase for even-boson (odd-boson) filling 
increases more (decreases more) for $U_2/U_0=0.4$ than for $U_2/U_0=0.2$. 
The convergence of the strong-coupling expansion  
from the first to the third order in $t$ is fairly good. 
We also find that higher-order terms mostly render the SF phase more stable
against the MI phase because the area of the MI phase 
mostly becomes smaller for higher-order expansions, which  
is similar to the case of spinless Bose--Hubbard model. 

\begin{figure}[htdp]
\includegraphics[height=3in]{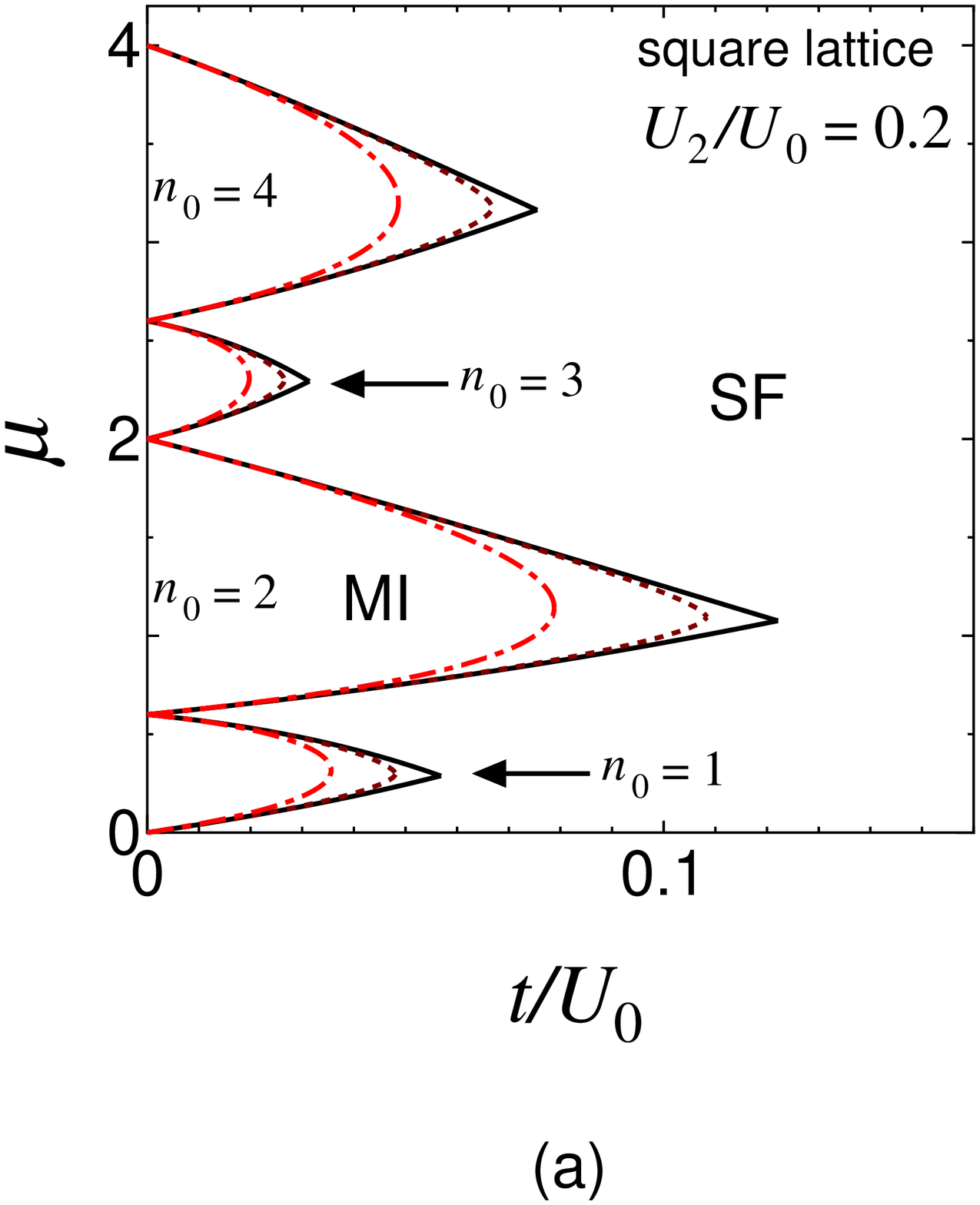}
\hspace{0.5in}
\includegraphics[height=3in]{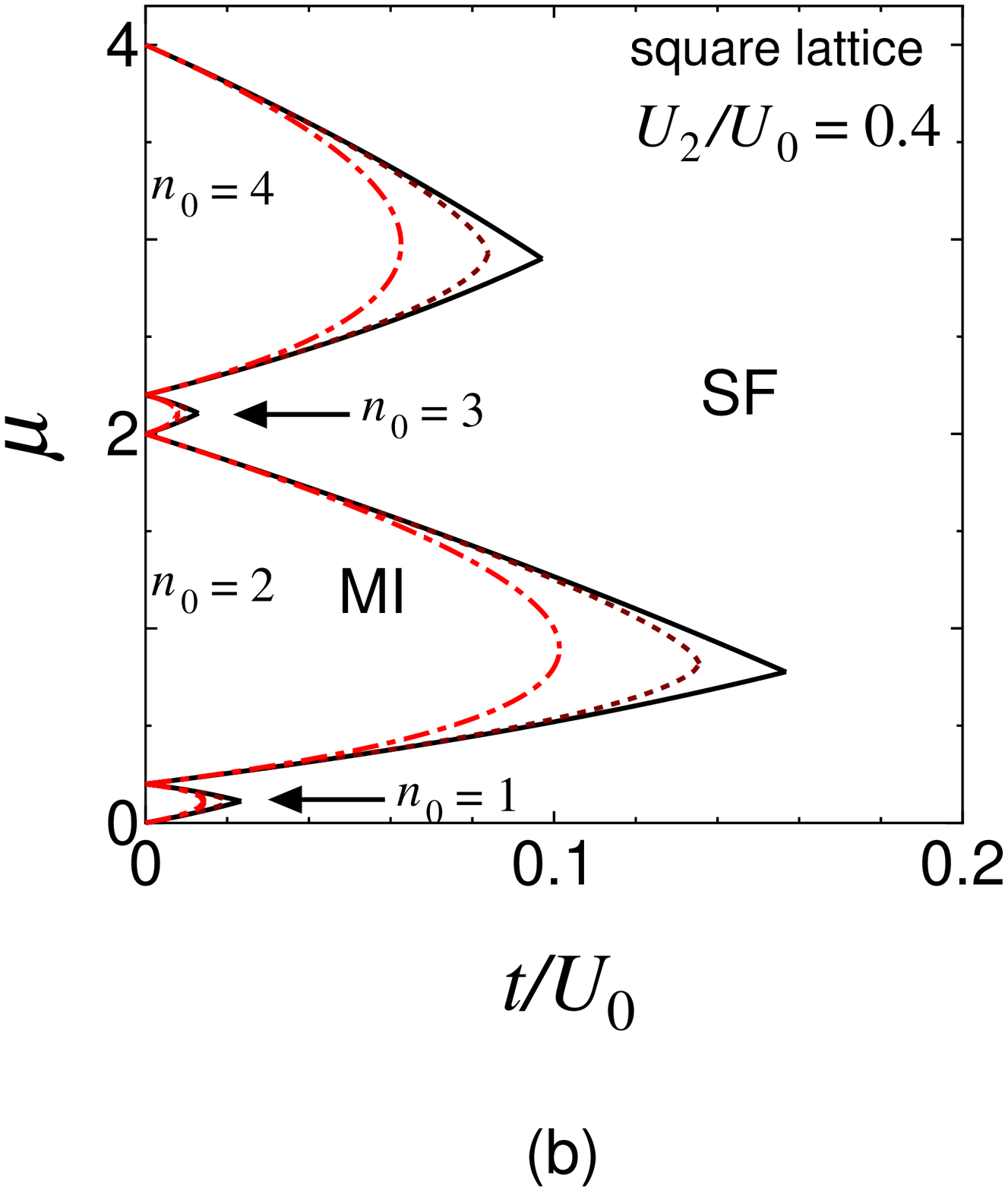}
\ \\ 
\caption{
(Color Online) Phase diagrams 
for (a) $U_2/U_0 = 0.2$ and for (b) $U_2/U_0 = 0.4$
obtained by the strong-coupling expansion 
(solid curve) and its chemical-potential fitting 
(brown dashed curve) in the square lattice.  
The red dot-dashed curve shows the PMFA results. 
}
\label{fit.sq}
\end{figure}

\begin{figure}[htdp]
\includegraphics[height=3in]{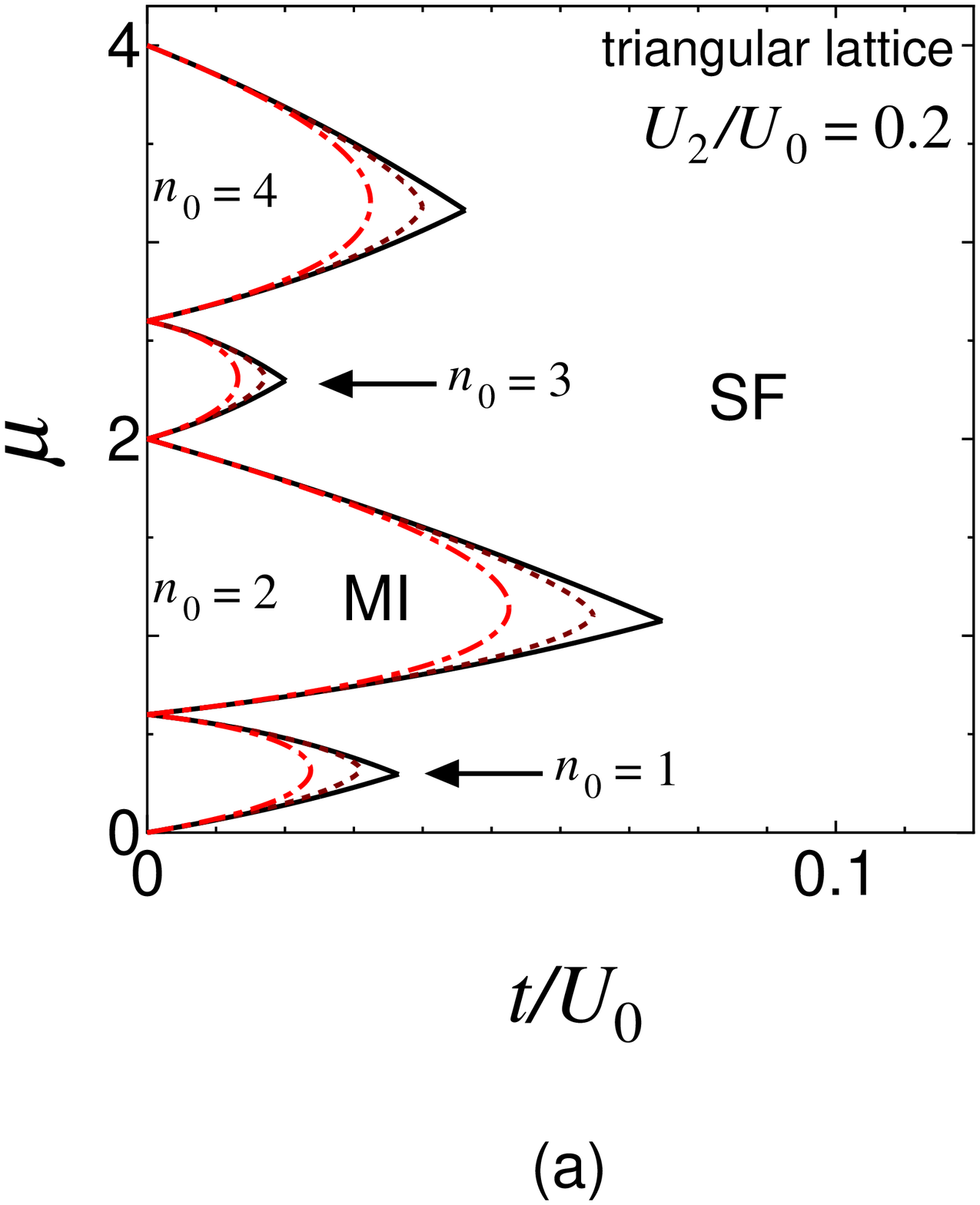}
\hspace{0.5in}
\includegraphics[height=3in]{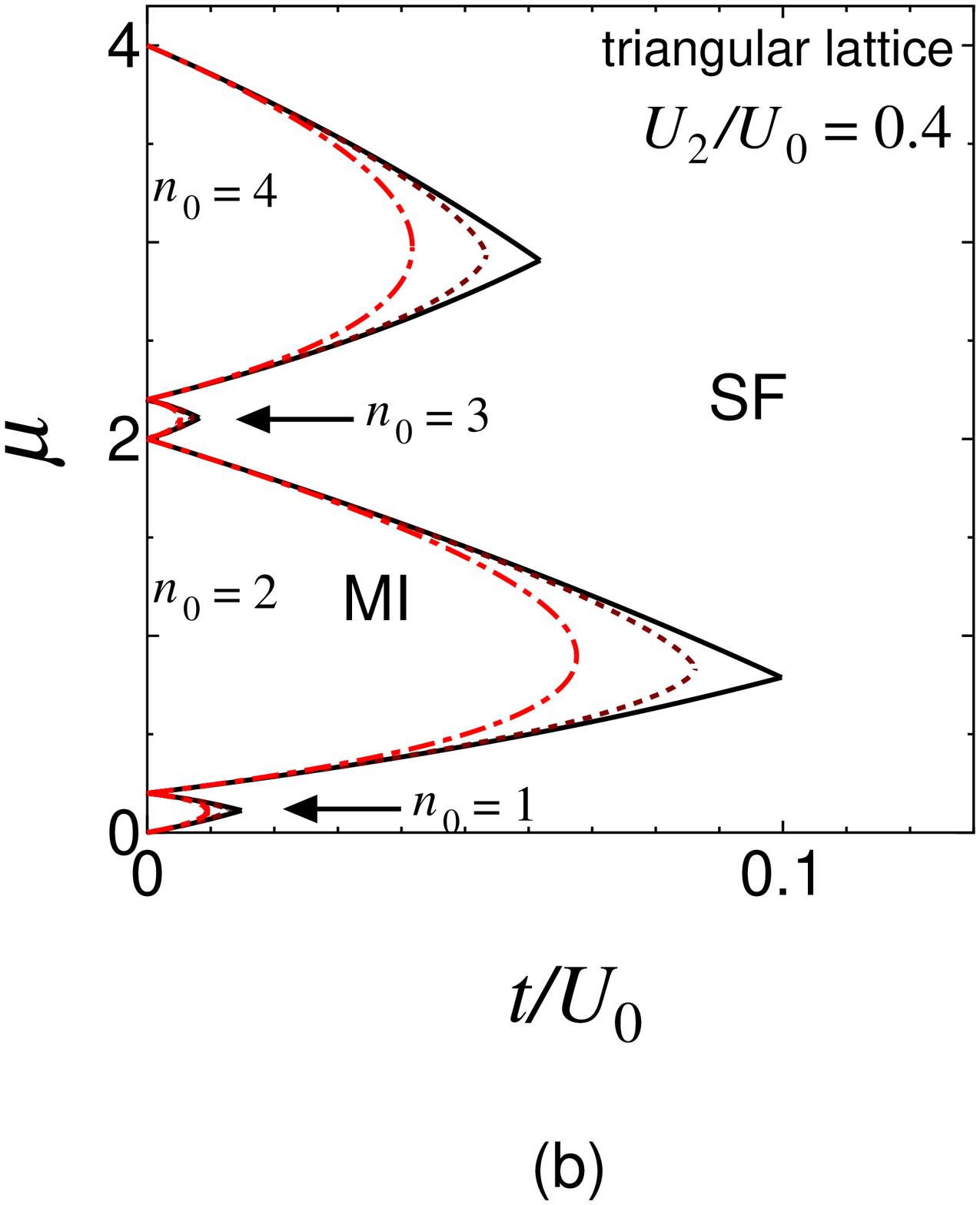}
\ \\ 
\caption{
(Color online) Same plot as in Fig. \ref{fit.sq}
but for (a) $U_2/U_0 = 0.2$ 
and (b) $U_2/U_0 = 0.2$ in the triangular lattice. 
}
\label{fit.tr}
\end{figure}

The phase-boundary curve obtained by 
the strong-coupling expansion up to the third order in $t$ 
has a cusp at the peak of the Mott lobe. 
However, the chemical potential in infinite order in $t$
should follow a power-law scaling near $t_{\rm C}$, 
which is the value of $t$ at the tip of the Mott lobe.  
Following Ref. \cite{Freericks} to the spinless Bose--Hubbard model in 2D, 
we perform a chemical-potential fitting method which assumes 
\begin{eqnarray}
\mu=A(t)\pm B(t)(t_{\rm C}-t)^{z\nu}. 
\end{eqnarray} 
with a critical exponent $z\nu\simeq 2/3$.  
Here $A(t)\approx a+bt+ct^2+dt^3$ and 
$B(t)\approx \alpha+\beta t+\gamma t^2$ 
are the regular polynomial functions of $t$. 
By using our strong-coupling expansion up to the third order
in $t$, we immediately determine $a$, $b$, $c$, and $d$ 
by setting $A(t)=[\mu^{\rm part}(t)+\mu^{\rm hole}(t)]/2$. 
In order to determine $\alpha$, $\beta$, $\gamma$, 
and $t_{\rm C}$ we compare the Taylor expansion of $t$ in 
$B(t)(t_{\rm C}-t)^{z\nu}$ with $[\mu^{\rm part}(t)-\mu^{\rm hole}(t)]/2$.  
Figures \ref{fit.sq} and \ref{fit.tr} 
compare the phase-boundary curves obtained by the 
third-order strong-coupling expansion
with those obtained by the chemical-potential fitting
in the square lattice and in the triangular lattice, respectively.  
The chemical-potential fitting 
makes the phase-boundary curves 
smooth and natural, as expected. 
The results obtained by the PMFA are also presented in both figures. 

For the square lattice, 
the QMC data show, for the $n_0=2$ Mott lobe, 
$t_{\rm C}/U_0\simeq 0.09(0.13)$ when $U_2/U_0=0.25(0.5)$
(as per our interpretation of Fig. 15 of Ref. \cite{QMC2D}). 
Our third-order strong-coupling results 
and the chemical-potential fitting show 
$t_{\rm C}/U_0=0.132(0.187)$ when $U_2/U_0=0.25(0.5)$ and
$t_{\rm C}/U_0=0.114(0.151)$ when $U_2/U_0=0.25(0.5)$, respectively. 
This shows that the chemical-potential fitting clearly improves the 
results obtained by the strong-coupling expansion up to third order in $t$. 
On the other hand, 
$t_{\rm C}/U_0=0.084(0.113)$ is obtained by the PMFA 
when $U_2/U_0=0.25(0.5)$. 
Hence the results obtained by QMC methods
are in between those obtained 
by the chemical-potential fitting and PMFA, 
at least in the case of the square lattice 
where QMC has been performed. 
This seems to reflect the fact that the PMFA includes 
a part of the infinite-order terms in $t$ but neglects the 
quantum fluctuations that possibly stabilize the MI phase, whereas 
the chemical-potential fitting based on the strong-coupling 
expansion up to third order in $t$ 
includes the quantum fluctuations but does not directly include 
the higher-order terms in $t$ that possibly stabilize the SF phase. 

Comparing the results obtained in the square lattice  
and those obtained in the triangular lattice,   
we find that $t_{\rm C}$ 
is roughly $6/4=3/2$ times greater in the square lattice 
than that in the triangular lattice, because of 
the difference between the number of nearest-neighbor 
sites $z$ [$z=4(6)$ in the square (triangular) lattice].   
However, there seems to be no significant qualitative difference 
between the results obtained in the square lattice 
and those obtained in the triangular lattice. 
These results suggest that there  the SF-MI transition has 
 no significant dependence on the lattice structure. 

\section{Summary}
In this paper, we present the strong-coupling expansion 
for the spin-1 Bose--Hubbard model 
for  the square and triangular lattices. 
In both lattices, we confirm that 
for filling with 
an even number of bosons, the MI phase 
 is considerably more stable
against the SF phase than  for filling with 
an odd number of bosons. We also fit the phase-boundary curves 
obtained by the strong-coupling expansion 
to the scaling form of the curves
by using the chemical-potential fitting method. 
The fitting curves smooth and natural as expected, and 
the values of $t_{\rm C}$ are clearly improved. 

Overall, there has been no essential difference 
in the phase boundary curves for 
the square and triangular lattices. 
This might be due to our assumption that the zeroth order in $t$
MI state is a nematic state. 
However, QMC study \cite{QMC2D} of the square lattice indicates 
that the magnetic structure factor shows no trace of magnetic order 
anywhere in the phase diagram. 
Thus, in contrast to the usual antiferromagnetic spin systems, 
there might be no significant dependence on the 
lattice structure for spin-1 bosons in two or more dimensions. 
\ack
The author acknowledges Makoto Yamashita and Yuta Toga 
for fruitful discussions about the spin-1 Bose--Hubbard model.

\subsection{Appendices}

\appendix
\section{Energies of Mott insulator and defect states in square and triangular lattices} 
By using $\Psi_{\rm even}$ 
and $\Psi_{\rm odd}$ [Eqs. (\ref{Me}) and (\ref{Mo})], the
energy per site of the MI state in the square lattice is 
\begin{eqnarray}
\frac{E_{\rm square,MI,even}(n_0)}{N}&=&\frac{U_0}{2}n_0(n_0-1)-U_2n_0-n_0\mu
-\frac{z_st^2}{3}\frac{n_0(n_0+3)}{U_0+2U_2}, 
\label{evenenergy}\\
\frac{E_{\rm square,MI,odd}(n_0)}{N}&=&
\frac{U_0}{2}n_0(n_0-1)-U_2(n_0-1)-n_0\mu\nonumber\\
&&-\frac{1}{9}z_st^2\Big[\frac{34}{25}\frac{(n_0+4)(n_0-1)}{U_0+4U_2}
+\frac{4}{5}\frac{2n_0^2+6n_0+7}{U_0+U_2}\nonumber\\
&&+\frac{(n_0+1)(n_0+2)}{U_0-2U_2}\Big]
\label{oddenergy}
\end{eqnarray}
up to the third order in $t$, and where $z_s=4$ is 
the number of nearest-neighbor sites
in the square lattice. 
The corresponding MI-state energy in the triangular lattice is 
\begin{eqnarray}
&&\frac{E_{\rm triangular,MI,even}(n_0)}{N}\nonumber\\
&=&\frac{U_0}{2}n_0(n_0-1)-U_2n_0-n_0\mu
-\frac{z_tt^2}{3}\frac{n_0(n_0+3)}{U_0+2U_2}
-\frac{2}{9}z_tt^3\frac{n_0(n_0+3)(2n_0+3)}{(U_0+2U_2)^2}, 
\label{evenenergytri}\\
&&\frac{E_{\rm triangular,MI,odd}(n_0)}{N}\nonumber\\
&=&\frac{U_0}{2}n_0(n_0-1)-U_2(n_0-1)-n_0\mu\nonumber\\
&&-\frac{1}{9}z_tt^2\Big[\frac{34}{25}\frac{(n_0-1)(n_0+4)}{U_0+4U_2}
+\frac{4}{5}\frac{2n_0^2+6n_0+7}{U_0+U_2}+\frac{(n_0+1)(n_0+2)}{U_0-2U_2}\Big]\nonumber\\
&&-\frac{2}{27}z_tt^3\Big[
\frac{118}{125}\frac{(n_0-1)(n_0+4)(2n_0+3)}{(U_0+4U_2)^2}
+\frac{32}{25}\frac{(n_0-1)(n_0+4)(2n_0+3)}{(U_0+4U_2)(U_0+U_2)}\nonumber\\
&&+\frac{4}{25}\frac{(n_0^2-1)(9n_0+1)+(n_0+2)(n_0+4)(9n_0+26)}{(U_0+U_2)^2}\nonumber\\
&&+\frac{8}{5}\frac{(n_0+1)(n_0+2)(2n_0+3)}{(U_0+U_2)(U_0-2U_2)}
+\frac{(n_0+1)(n_0+2)(2n_0+3)}{(U_0-2U_2)^2}\Big]
\label{oddenergytri}
\end{eqnarray}
up to the third order in $t$, and where 
$z_t=6$ is the number of nearest-neighbor sites  
in the triangular lattice. Up to the second order in $t$, 
the expressions for the MI energy is the same for both lattices 
except for the difference between $z_s$ and $z_t$. However, 
in contrast to the square lattice, the third-order terms in $t$ exist 
for the triangular lattice. 

The energies of the defect states can be obtained 
by using $\Psi^{\rm part}_{\rm even}$,
$\Psi^{\rm hole}_{\rm even}$, 
$\Psi^{\rm part}_{\rm odd}$, and 
$\Psi^{\rm hole}_{\rm odd}$ [Eqs. (\ref{pe})--(\ref{ho})]
up to the third order in $t$. 
For the square lattice, we obtain
\begin{eqnarray}
&&E^{\rm part}_{\rm square,def,even}(n_0)-E_{\rm square,MI,even}(n_0)
\nonumber\\
&=& U_0n-\mu-z_st\frac{n_0+3}{3}\nonumber\\
&& -\frac{z_s(z_s-7)t^2}{9}\frac{n_0(n_0+3)}{U_0+2U_2}
-\frac{z_st^2n_0}{9}\Big[2\Big(\frac{n_0+5}{2U_0+3U_2}
            +\frac{n_0+3}{3U_2}\Big)
            +\frac{n_0+2}{2U_0}\Big]\nonumber\\
&&-\frac{z_st^3}{27}n_0(n_0+3)
        \Big\{
        (z_s-1)\Big[
           \frac{(2n_0+3)z_s-3(3n_0+8)}{(U_0+2U_2)^2}\nonumber\\
&&           +\frac{2}{U_0+2U_2}\Big(
           2\frac{n_0+5}{2U_0+3U_2}+\frac{n_0+2}{2U_0}\Big)
           +\frac{4(n_0+3)}{3U_2(U_0+2U_2)}\Big]\nonumber\\
&&        -z_s\Big[
          2\Big(\frac{n_0+5}{(2U_0+3U_2)^2}+\frac{n_0+3}{(3U_2)^2}\Big)
          +\frac{n_0+2}{(2U_0)^2}\Big]
        +\frac{4}{3U_2}\Big(
        \frac{1}{5}\frac{n_0+5}{2U_0+3U_2}+\frac{n_0+2}{2U_0}\Big)
        \Big\},
\label{evenpart}
\end{eqnarray}
\begin{eqnarray}
&&E^{\rm hole}_{\rm square,def,even}(n_0)-E_{\rm square,MI,even}(n_0)\nonumber\\
&=&-U_0(n_0-1)+2U_2+\mu-z_st\frac{n_0}{3}\nonumber\\
&&-\frac{z_s(z_s-7)t^2}{9}\frac{n_0(n_0+3)}{U_0+2U_2}
-\frac{z_st^2(n_0+3)}{9}\Big[2\Big(\frac{n_0-2}{2U_0+3U_2}
            +\frac{n_0}{3U_2}\Big)
            +\frac{n_0+1}{2U_0}\Big]\nonumber\\
&&-\frac{z_st^3}{27}n_0(n_0+3)
        \Big\{
        (z_s-1)\Big[
           \frac{(2n_0+3)z_s-3(3n_0+1)}{(U_0+2U_2)^2}\nonumber\\
&&           +\frac{2}{U_0+2U_2}\Big(
           2\frac{n_0-2}{2U_0+3U_2}+\frac{n_0+1}{2U_0}\Big)
           +\frac{4n_0}{3U_2(U_0+2U_2)}\Big]\nonumber\\
&&        -z_s\Big[
          2\Big(\frac{n_0-2}{(2U_0+3U_2)^2}+\frac{n_0}{(3U_2)^2}\Big)
          +\frac{n_0+1}{(2U_0)^2}\Big]
        +\frac{4}{3U_2}\Big(
        \frac{1}{5}\frac{n_0-2}{2U_0+3U_2}+\frac{n_0+1}{2U_0}\Big)
        \Big\},\label{evenhole}
\end{eqnarray}
\begin{eqnarray}
&&E^{\rm part}_{\rm square,def,odd}(n_0)-E_{\rm square,MI,odd}(n_0)\nonumber\\
&=&U_0n_0-2U_2-z_st\frac{n_0+1}{3}-\mu\nonumber\\
&&-\frac{z_s(z_s-3)t^2}{9}(n_0+1)\Big[\frac{n_0+2}{U_0-2U_2}
          +\frac{4}{5}\frac{n_0-1}{U_0+U_2}\Big]\nonumber\\
&&-\frac{z_st^2}{9}(n_0+4)\Big[2\frac{n_0-1}{2U_0+3U_2}
                    +\frac{n_0+2}{2U_0}
-\frac{68}{25}\frac{n_0-1}{U_0+4U_2}
-\frac{8}{5}\frac{n_0+2}{U_0+U_2}\Big]\nonumber\\
&&-\frac{2}{45}z_s(2z_s+3)t^2\frac{(n_0+1)(n_0+4)}{3U_2}\nonumber\\
&&-\frac{z_s(z_s-1)t^3}{27}\frac{(n_0+1)(n_0+2)}{(U_0-2U_2)^2}
                     \big[(2n_0+3)z_s-(5n_0+6)\big]\nonumber\\
&&-\frac{4}{675}z_s(z_s-1)t^3\frac{n_0+1}{(U_0+U_2)^2}
                     \big[(n_0-1)(9n_0+1)z_s-2(17n_0^2+26n_0+32)\big]\nonumber\\
&&-\frac{z_s(z_s-1)^2t^3}{27}\frac{n_0+1}{U_0+U_2}
\Big[\frac{32}{25}\frac{(n_0-1)(n_0+4)}{U_0+4U_2}
     +\frac{8}{5}\frac{(n_0+2)(2n_0+3)}{U_0-2U_2}\Big]\nonumber\\
&&-\frac{2}{27}z_s(z_s-1)t^3(n_0+1)(n_0+4)
\Big\{\frac{n_0-1}{2U_0+3U_2}\Big[\frac{34}{25}\frac{1}{U_0+4U_2}
             +\frac{4}{5}\frac{1}{U_0+U_2}\Big]\nonumber\\
&&      +\frac{n_0+2}{2U_0}\Big[\frac{1}{U_0-2U_2}
             +\frac{4}{5}\frac{1}{U_0+U_2}\Big]
      +\frac{1}{3U_2}\Big[\frac{2}{25}(8z_s+9)\frac{n_0-1}{U_0+4U_2}
             +\frac{4}{5}z_s\frac{n_0+2}{U_0+U_2}\Big]\Big\}\nonumber\\
&&-\frac{4}{135}z_s(2z_s+3)t^3\frac{(n_0+1)(n_0+4)}{3U_2}
      \Big[\frac{1}{5}\frac{n_0-1}{2U_0+3U_2}+\frac{n_0+2}{2U_0}\Big]\nonumber\\
&&+\frac{2}{675}z_st^3\frac{(n_0+1)(n_0+4)}{(3U_2)^2}
      \big[2(n_0-11)z_s^2+9(3n_0+7)z_s-9(n_0+4)\big]\nonumber\\
&&+\frac{z_st^3}{27}(n_0+1)(n_0+4)\Big[
        \frac{68}{25}(z_s-1)\frac{n_0-1}{(U_0+4U_2)^2}\nonumber\\
        &&+2z_s\frac{n_0-1}{(2U_0+3U_2)^2}+z_s\frac{n_0+2}{(2U_0)^2}\Big],
\label{oddpart}
\end{eqnarray}
\begin{eqnarray}
&&E^{\rm hole}_{\rm square,def,odd}(n_0)-E_{\rm square,MI,odd}(n_0)\nonumber\\
&=&-U_0(n_0-1)-z_st\frac{n_0+2}{3}+\mu\nonumber\\
&&-\frac{z_s(z_s-3)t^2}{9}(n_0+2)\Big[\frac{n_0+1}{U_0-2U_2}
          +\frac{4}{5}\frac{n_0+4}{U_0+U_2}\Big]\nonumber\\
&&-\frac{z_st^2}{9}(n_0-1)\Big[2\frac{n_0+4}{2U_0+3U_2}
                    +\frac{n_0+1}{2U_0}
-\frac{68}{25}\frac{n_0+4}{U_0+4U_2}
-\frac{8}{5}\frac{n_0+1}{U_0+U_2}\Big]\nonumber\\
&&-\frac{2}{45}z_s(2z_s+3)t^2\frac{(n_0-1)(n_0+2)}{3U_2}\nonumber\\
&&-\frac{z_s(z_s-1)t^3}{27}\frac{(n_0+1)(n_0+2)}{(U_0-2U_2)^2}
                     \big[(2n_0+3)z_s-(5n_0+9)\big]\nonumber\\
&&-\frac{4}{675}z_s(z_s-1)t^3\frac{n_0+2}{(U_0+U_2)^2}
                     \big[(n_0+4)(9n_0+26)z_s-2(17n_0^2+76n_0+107)\big]\nonumber\\
&&-\frac{z_s(z_s-1)^2t^3}{27}\frac{n_0+2}{U_0+U_2}
\Big[\frac{32}{25}\frac{(n_0-1)(n_0+4)}{U_0+4U_2}
     +\frac{8}{5}\frac{(n_0+1)(2n_0+3)}{U_0-2U_2}\Big]\nonumber\\
&&-\frac{2}{27}z_s(z_s-1)t^3(n_0-1)(n_0+2)
\Big\{\frac{n_0+4}{2U_0+3U_2}\Big[\frac{34}{25}\frac{1}{U_0+4U_2}
             +\frac{4}{5}\frac{1}{U_0+U_2}\Big]\nonumber\\
&&      +\frac{n_0+1}{2U_0}\Big[\frac{1}{U_0-2U_2}
             +\frac{4}{5}\frac{1}{U_0+U_2}\Big]
      +\frac{1}{3U_2}\Big[\frac{2}{25}(8z_s+9)\frac{n_0+4}{U_0+4U_2}
             +\frac{4}{5}z_s\frac{n_0+1}{U_0+U_2}\Big]\Big\}\nonumber\\
&&-\frac{4}{135}z_s(2z_s+3)t^3\frac{(n_0-1)(n_0+2)}{3U_2}
      \Big[\frac{1}{5}\frac{n_0+4}{2U_0+3U_2}+\frac{n_0+1}{2U_0}\Big]\nonumber\\
&&+\frac{2}{675}z_st^3\frac{(n_0-1)(n_0+2)}{(3U_2)^2}
      \big[2(n_0+14)z_s^2+9(3n_0+2)z_s-9(n_0-1)\big]\nonumber\\
&&+\frac{z_st^3}{27}(n_0-1)(n_0+2)\Big[
        \frac{68}{25}(z_s-1)\frac{n_0+4}{(U_0+4U_2)^2}\nonumber\\
        &&+2z_s\frac{n_0+4}{(2U_0+3U_2)^2}+z_s\frac{n_0+1}{(2U_0)^2}\Big]. 
\label{oddhole}
\end{eqnarray}
For the triangular lattice, we obtain
\begin{eqnarray}
&&E^{\rm part}_{\rm triangular,def,even}(n_0)-E_{\rm triangular,MI,even}(n_0)\nonumber\\
&=& U_0n-\mu-z_tt\frac{n_0+3}{3}\nonumber\\
&& -\frac{z_t(z_t-7)t^2}{9}\frac{n_0(n_0+3)}{U_0+2U_2}
-\frac{z_tt^2n_0}{9}\Big[2\Big(\frac{n_0+5}{2U_0+3U_2}
            +\frac{n_0+3}{3U_2}\Big)
            +\frac{n_0+2}{2U_0}\Big]\nonumber\\
&&-\frac{z_tt^3}{27}n_0\Big\{
  \frac{n_0+3}{(U_0+2U_2)^2}
\big[(2n_0+3)z_t^2-2(4n_0+9)z_t-3(17n_0+32)\big]\nonumber\\
&& -2\frac{n_0+5}{(2U_0+3U_2)^2}\big[(n_0+3)z_t-2n_0\big]
-\frac{n_0+2}{(2U_0)^2}\big[(n_0+3)z_t-2n_0\big]
  -2(z_t-2)\frac{(n_0+3)^2}{(3U_2)^2}\nonumber\\
&&+2(z_t+1)\frac{n_0+3}{U_0+2U_2}
  \Big[2\frac{n_0+5}{2U_0+3U_2}+\frac{n_0+2}{2U_0}\Big]\nonumber\\
&&+4\frac{n_0+3}{3U_2}\Big[\frac{(n_0+3)z+n_0-3}{U_0+2U_2}
+\frac{1}{5}\frac{n_0+5}{2U_0+3U_2}+\frac{n_0+2}{2U_0}\Big]\Big\}
\label{evenparttri}
\end{eqnarray}
\begin{eqnarray}
&&E^{\rm hole}_{\rm triangular,def,even}(n_0)-E_{\rm triangular,MI,even}(n_0)\nonumber\\
&=&-U_0(n_0-1)+2U_2+\mu-z_tt\frac{n_0}{3}\nonumber\\
&&-\frac{z_t(z_t-7)t^2}{9}\frac{n_0(n_0+3)}{U_0+2U_2}
-\frac{z_tt^2(n_0+3)}{9}\Big[2\Big(\frac{n_0-2}{2U_0+3U_2}
            +\frac{n_0}{3U_2}\Big)
            +\frac{n_0+1}{2U_0}\Big]\nonumber\\
&&-\frac{z_tt^3}{27}(n_0+3)\Big\{
  \frac{n_0}{(U_0+2U_2)^2}
\big[(2n_0+3)z_t^2-2(4n_0+3)z_t-3(17n_0+19)\big]\nonumber\\
&& -2\frac{n_0-2}{(2U_0+3U_2)^2}\big[n_0z_t-2(n_0+3)\big]
-\frac{n_0+1}{(2U_0)^2}\big[n_0z_t-2(n_0+3)\big]
  -2(z_t-2)\frac{n_0^2}{(3U_2)^2}\nonumber\\
&&+2(z_t+1)\frac{n_0}{U_0+2U_2}
  \Big[2\frac{n_0-2}{2U_0+3U_2}+\frac{n_0+1}{2U_0}\Big]\nonumber\\
&&+4\frac{n_0}{3U_2}\Big[\frac{n_0z+n_0+6}{U_0+2U_2}
+\frac{1}{5}\frac{n_0-2}{2U_0+3U_2}+\frac{n_0+1}{2U_0}\Big]\Big\}
\label{evenholetri}
\end{eqnarray}
\begin{eqnarray}
&&E^{\rm part}_{\rm triangular,def,odd}(n_0)-E_{\rm triangular,MI,odd}(n_0)\nonumber\\
&=&U_0n_0-2U_2-z_tt\frac{n_0+1}{3}-\mu\nonumber\\
&&-\frac{z_t(z_t-3)t^2}{9}(n_0+1)\Big[\frac{n_0+2}{U_0-2U_2}
          +\frac{4}{5}\frac{n_0-1}{U_0+U_2}\Big]\nonumber\\
&&-\frac{z_tt^2}{9}(n_0+4)\Big[2\frac{n_0-1}{2U_0+3U_2}
                    +\frac{n_0+2}{2U_0}
-\frac{68}{25}\frac{n_0-1}{U_0+4U_2}
-\frac{8}{5}\frac{n_0+2}{U_0+U_2}\Big]\nonumber\\
&&-\frac{2}{45}z_t(2z_t+3)t^2\frac{(n_0+1)(n_0+4)}{3U_2}\nonumber\\
&&+\frac{2z_tt^3}{3375}\frac{(n_0-1)(n_0+4)}{(U_0+4U_2)^2}
\big[85(n_0+1)z_t+2(524n_0+701)\big]\nonumber\\
&&+\frac{4}{675}\frac{z_tt^3}{(U_0+U_2)^2}
\Big\{(n_0^2-1)\big[-(9n_0+1)z_t^2+4(7n_0+3)z_t+8(11n_0+4)\big]\nonumber\\
&&+(n_0+2)(n_0+4)\big[5(n_0+1)z_t+2(37n_0+88)\big]\Big\} \nonumber\\
&&+\frac{z_tt^3}{27}\frac{(n_0+1)(n_0+2)}{(U_0-2U_2)^2}
\big[-(2n_0+3)z_t^2+2(3n_0+4)z_t+19n_0+26\big]\nonumber\\
&&+\frac{2z_tt^3}{675}\frac{(n_0-1)(n_0+4)}{(2U_0+3U_2)^2}
\big[25(n_0+1)z_t-34(n_0-1)\big]\nonumber\\
&&+\frac{z_tt^3}{27}\frac{(n_0+2)(n_0+4)}{(2U_0)^2}
\big[(n_0+1)z_t-2(n_0+2)\big]\nonumber\\
&&+\frac{2z_tt^3}{675}\frac{(n_0+1)(n_0+4)}{(3U_2)^2}
\big[2(n_0-11)z_t^2+9(3n_0+7)z_t-27(n_0+4)\big]\nonumber\\
&&-\frac{32z_tt^3}{675}\frac{(n_0-1)(n_0+4)}{(U_0+4U_2)(U_0+U_2)}
\big[(n_0+1)z_t^2-2(n_0+1)z_t-(13n_0+19)\big]\nonumber\\
&&-\frac{8}{135}z_t(z_t^2-2z_t-7)t^3\frac{(n_0+1)(n_0+2)(2n_0+3)}
{(U_0+U_2)(U_0-2U_2)}-\frac{16z_tt^3}{135}\frac{(n_0-1)(n_0+2)(n_0+4)}{2U_0(2U_0+3U_2)}\nonumber\\
&&-\frac{68z_tt^3}{675}\frac{(n_0-1)(n_0+4)}{(2U_0+3U_2)(U_0+4U_2)}
\big[(n_0+1)z_t+n_0+7\big]\nonumber\\
&&-\frac{8}{135}z_t(z_t+1)t^3\frac{(n_0^2-1)(n_0+4)}{(2U_0+3U_2)(U_0+U_2)}
-\frac{8z_tt^3}{135}\frac{(n_0+2)(n_0+4)}{2U_0(U_0+U_2)}
\big[(n_0+1)z_t+n_0+7\big]\nonumber\\
&&-\frac{2}{27}z_t(z_t+1)t^3\frac{(n_0+1)(n_0+2)(n_0+4)}{2U_0(U_0-2U_2)}
-\frac{4}{675}z_t(8z_t^2+z_t+9)t^3\frac{(n_0^2-1)(n_0+4)}{3U_2(U_0+4U_2)}\nonumber\\
&&-\frac{8}{135}z_t^2(z_t-1)t^3\frac{(n_0+1)(n_0+2)(n_0+4)}{3U_2(U_0+U_2)}
-\frac{4}{675}z_t(2z_t+3)t^3\frac{(n_0^2-1)(n_0+4)}{3U_2(2U_0+3U_2)}\nonumber\\
&&-\frac{4}{135}z_t(2z_t+3)t^3\frac{(n_0+1)(n_0+2)(n_0+4)}{3U_2\cdot2U_0}
\label{oddparttri}
\end{eqnarray}
\begin{eqnarray}
&&E^{\rm hole}_{\rm triangular,def,odd}(n_0)-E_{\rm triangular,MI,odd}(n_0)\nonumber\\
&=&-U_0(n_0-1)-z_tt\frac{n_0+2}{3}+\mu\nonumber\\
&&-\frac{z_t(z_t-3)t^2}{9}(n_0+2)\Big[\frac{n_0+1}{U_0-2U_2}
          +\frac{4}{5}\frac{n_0+4}{U_0+U_2}\Big]\nonumber\\
&&-\frac{z_tt^2}{9}(n_0-1)\Big[2\frac{n_0+4}{2U_0+3U_2}
                    +\frac{n_0+1}{2U_0}
-\frac{68}{25}\frac{n_0+4}{U_0+4U_2}
-\frac{8}{5}\frac{n_0+1}{U_0+U_2}\Big]\nonumber\\
&&-\frac{2}{45}z_t(2z_t+3)t^2\frac{(n_0-1)(n_0+2)}{3U_2}\nonumber\\
&&+\frac{2z_tt^3}{3375}\frac{(n_0-1)(n_0+4)}{(U_0+4U_2)^2}
\big[85(n_0+2)z_t+2(524n_0+871)\big]\nonumber\\
&&+\frac{4}{675}\frac{z_tt^3}{(U_0+U_2)^2}
\Big\{(n_0+2)(n_0+4)\big[-(9n_0+26)z_t^2+4(7n_0+18)z_t+8(11n_0+29)\big]\nonumber\\
&&+(n_0^2-1)\big[5(n_0+2)z_t+2(37n_0+23)\big]\Big\} \nonumber\\
&&+\frac{z_tt^3}{27}\frac{(n_0+1)(n_0+2)}{(U_0-2U_2)^2}
\big[-(2n_0+3)z_t^2+2(3n_0+5)z_t+19n_0+31\big]\nonumber\\
&&+\frac{2z_tt^3}{675}\frac{(n_0-1)(n_0+4)}{(2U_0+3U_2)^2}
\big[25(n_0+2)z_t-34(n_0+4)\big]\nonumber\\
&&+\frac{z_tt^3}{27}\frac{n_0^2-1}{(2U_0)^2}
\big[(n_0+2)z_t-2(n_0+1)\big]\nonumber\\
&&+\frac{2z_tt^3}{675}\frac{(n_0-1)(n_0+2)}{(3U_2)^2}
\big[2(n_0+14)z_t^2+9(3n_0+2)z_t-27(n_0-1)\big]\nonumber\\
&&-\frac{32z_tt^3}{675}\frac{(n_0-1)(n_0+4)}{(U_0+4U_2)(U_0+U_2)}
\big[(n_0+2)z_t^2-2(n_0+2)z_t-(13n_0+20)\big]\nonumber\\
&&-\frac{8}{135}z_t(z_t^2-2z_t-7)t^3\frac{(n_0+1)(n_0+2)(2n_0+3)}
{(U_0+U_2)(U_0-2U_2)}-\frac{16z_tt^3}{135}\frac{(n_0^2-1)(n_0+4)}{2U_0(2U_0+3U_2)}\nonumber\\
&&-\frac{68z_tt^3}{675}\frac{(n_0-1)(n_0+4)}{(2U_0+3U_2)(U_0+4U_2)}
\big[(n_0+2)z_t+n_0-4\big]\nonumber\\
&&-\frac{8}{135}z_t(z_t+1)t^3\frac{(n_0-1)(n_0+2)(n_0+4)}{(2U_0+3U_2)(U_0+U_2)}
-\frac{8z_tt^3}{135}\frac{n_0^2-1}{2U_0(U_0+U_2)}
\big[(n_0+2)z_t+n_0-4\big]\nonumber\\
&&-\frac{2}{27}z_t(z_t+1)t^3\frac{(n_0^2-1)(n_0+2)}{2U_0(U_0-2U_2)}
-\frac{4}{675}z_t(8z_t^2+z_t+9)t^3
\frac{(n_0-1)(n_0+2)(n_0+4)}{3U_2(U_0+4U_2)}\nonumber\\
&&-\frac{8}{135}z_t^2(z_t-1)t^3
\frac{(n_0^2-1)(n_0+2)}{3U_2(U_0+U_2)}
-\frac{4}{675}z_t(2z_t+3)t^3
\frac{(n_0-1)(n_0+2)(n_0+4)}{3U_2(2U_0+3U_2)}\nonumber\\
&&-\frac{4}{135}z_t(2z_t+3)t^3\frac{(n_0^2-1)(n_0+2)}{3U_2\cdot2U_0}
\label{oddholetri}
\end{eqnarray}

As for the MI energy, 
the expressions for the MI energy up to the second order in $t$ 
are the same for both lattices 
except for the difference between $z_s$ and $z_t$ but 
the third-order terms in $t$ are different from each lattice. 
This difference between the third-order terms originates from 
 particles or holes that can (cannot) return to their 
original site through three hopping processes 
in the triangular (square) lattice. 
%By equating the right-hand side of Eqs. (\ref{evenpart})--(\ref{oddhole})
%with z_sero, we obtain the SF--MI phase-boundary curve.  

\end{document}